\documentclass[runningheads]{llncs}

\usepackage{multirow}
\usepackage[T1]{fontenc}
\usepackage{bm}
\usepackage{adjustbox}
\usepackage{graphicx}
\begin{document}
\title{APNet2: High-quality and High-efficiency Neural Vocoder
	with Direct Prediction of Amplitude and
	Phase Spectra\thanks{This work was funded by the Anhui Provincial Natural Science Foundation under Grant 2308085QF200 and the Fundamental Research
		Funds for the Central Universities under Grant WK2100000033.}}
\titlerunning{APNet2}
%
\author{Hui-Peng Du \and Ye-Xin Lu \and Yang Ai \and Zhen-Hua Ling}
\authorrunning{H.-P. Du et al.}
%
\institute{National Engineering Research Center of Speech and Language Information Processing, \\University of Science and Technology of China, Hefei, P. R. China \\
	\email{\{redmist, yxlu0102\}@mail.ustc.edu.cn},  \email{\{yangai, zhling\}@ustc.edu.cn}}
\maketitle              
\begin{abstract}
In our previous work, we proposed a neural vocoder called APNet, which directly predicts speech amplitude and phase spectra with a 5 ms frame shift in parallel from the input acoustic features, and then reconstructs the 16 kHz speech waveform using inverse short-time Fourier transform (ISTFT). 
APNet demonstrates the capability to generate synthesized speech of comparable quality to the HiFi-GAN vocoder but with a considerably improved inference speed.
However, the performance of the APNet vocoder is constrained by the waveform sampling rate and spectral frame shift, limiting its practicality for high-quality speech synthesis.
Therefore, this paper proposes an improved iteration of APNet, named APNet2. 
The proposed APNet2 vocoder adopts ConvNeXt v2 as the backbone network for amplitude and phase predictions, expecting to enhance the modeling capability.
Additionally, we introduce a multi-resolution discriminator (MRD) into the GAN-based losses and optimize the form of certain losses.
At a common configuration with a waveform sampling rate of 22.05 kHz and spectral frame shift of 256 points (i.e., approximately 11.6ms), our proposed APNet2 vocoder outperformed the original APNet and Vocos vocoders in terms of synthesized speech quality. 
The synthesized speech quality of APNet2 is also comparable to that of HiFi-GAN and iSTFTNet, while offering a significantly faster inference speed.

\keywords{Neural vocoder \and Amplitude spectrum \and Phase spectrum \and ConvNeXt v2 \and Multi-resolution discriminator.}
\end{abstract}

\section{Introduction}

Neural vocoder technology, which converts speech acoustic features into waveforms, has seen rapid progress in recent years.
The vocoder capability significantly affects the  performance of several speech-generation applications, such as text-to-speech (TTS) synthesis, singing voice synthesis (SVS), bandwidth extension (BWE), speech enhancement (SE), and voice conversion (VC).

The synthesized speech quality and inference efficiency are two major indicators for evaluating vocoders.
Although auto-regressive (AR) neural vocoders such as WaveNet \cite{oord2016wavenet} and SampleRNN \cite{mehri2016samplernn} achieved significant improvements in synthesized speech quality compared to traditional signal-processing-based vocoders \cite{kawahara1999restructuring,morise2016world}, they demanded considerable computational cost and exhibited extremely low generation efficiency due to their auto-regressive inference mode on raw waveforms.
Consequently, alternative approaches have been proposed, including knowledge-distilling-based models (e.g., Parallel WaveNet \cite{oord2018parallel} and ClariNet \cite{ping2018clarinet}), flow-based models (e.g., WaveGlow \cite{prenger2019waveglow} and WaveFlow \cite{ping2020waveflow}), and glottis-based models (e.g., GlotNet \cite{juvela2019glotnet} and LPCNet \cite{valin2019lpcnet}).
While these models have substantially improved inference efficiency, their overall computational complexity remains elevated, limiting their applicability in resource-constrained environments like embedded devices.
Recently, there has been increasing attention towards waveform generation models that eschew auto-regressive or flow-like structures.
One such model is the neural source-filter (NSF) model \cite{wang2019neural}, which combines speech production mechanisms with neural networks to predict speech waveforms directly based on explicit F0 and mel-spectrograms.
Additionally, generative adversarial network (GAN) \cite{goodfellow2014generative} based vocoders, including WaveGAN \cite{donahue2018adversarial}, MelGAN \cite{kumar2019melgan}, and HiFi-GAN \cite{kong2020hifi}, leverage GANs to ensure high-quality synthesized speech while simultaneously enhancing efficiency.
These methods directly predict waveforms without relying on complex structures.
However, to compensate for the difference in temporal resolution between input acoustic features and output waveforms, typical time-domain GAN vocoders need multiple transposed convolutions to upsample the input features to the desired sample rate, incurring substantial computational cost.

To solve this issue, several vocoders (e.g., iSTFTNet \cite{kaneko2022iSTFTNet}, Vocos \cite{siuzdak2023vocos}, and APNet \cite{ai2023apnet}) turn to predict amplitude and phase spectra and finally use inverse short-time Fourier transform (ISTFT) to reconstruct waveforms.
This type of approach effectively avoids the direct prediction of high-resolution waveforms.
Compared with HiFi-GAN \cite{kong2020hifi}, iSTFTNet \cite{kaneko2022iSTFTNet} employs fewer upsampling layers to estimate amplitude and phase spectra at a large time-domain resolution, which still consumes much computational cost.
Furthermore, Vocos \cite{siuzdak2023vocos} leverages ConvNeXt \cite{liu2022convnet} as its backbone network, omitting upsampling layers to directly predict amplitude and phase spectra at the same temporal resolution as the input acoustic features.
However, without effective tools for precise phase estimation, iSTFTNet and Vocos only define waveform- and amplitude-related losses without constraining the predicted phase spectra.
This may lead to insufficient phase prediction accuracy and black-box problem of phase prediction.
In our previous work, we proposed the APNet vocoder \cite{ai2023apnet} as a means to overcome this limitation.
The APNet vocoder can also directly predict the amplitude and phase spectra at the original resolution and then reconstruct the waveforms.
Differently, to ensure phase prediction accuracy, the APNet vocoder adopts a phase parallel estimation architecture and employs anti-wrapping losses to explicitly model and optimize the phase, respectively.
Through experiments, the APNet vocoder achieved comparable synthesized speech quality as the HiFi-GAN vocoder, while achieving a remarkable eightfold increase in inference speed on a CPU.

However, in further investigation, we discover that phase prediction is highly sensitive to frame shift \cite{ai2023long}.
When frame shift increases, phase continuity deteriorates, and the modeling capabilities of the existing APNet vocoder may prove insufficient.
In preliminary experiments, we draw a conclusion that accurately modeling the temporal discontinuity of long-frame-shift phase spectra poses challenges for the APNet vocoder, which can lead to a noticeable reduction in phase prediction accuracy.
But, accurately predicting long-frame-shift spectra is crucial in several speech generation tasks.
One particular challenge arises when generating short-frame-shift features using attention-based acoustic models \cite{shen2018natural,wang2017tacotron}, as aligning texts and features can be difficult and result in improper alignment.
Besides, a sampling rate of 16 kHz may not meet the requirements for high-quality speech synthesis applications.
Currently, most neural vocoders take 80-dim mel-spectrograms in the 0$\sim$8000 Hz frequency range as input and generate speech waveforms at a 22.05 kHz sampling rate for a fair comparison.
This process also implicitly includes the operation of BWE.

Therefore, this paper proposes an improved iteration of APNet called APNet2.
The proposed APNet2 vocoder further  augments its modeling capabilities, making it adaptable to scenarios with higher sampling rates and extended frame shifts.
Similar to APNet, APNet2 is also composed of an amplitude spectrum predictor (ASP) and a phase spectrum predictor (PSP), which predict the speech amplitude and phase spectra from acoustic features in parallel, and then reconstruct the waveforms through ISTFT.
Differently, the APNet2 vocoder adopts the ConvNeXt v2 \cite{woo2023convnext} as the backbone network of both ASP and PSP, bolstering its modeling capabilities.
Regarding the phase-related loss function, we have updated the anti-wrapping function from a negative cosine function to a linear function to yield more precise phase predictions.
In the waveform-related GAN loss, we introduce a multi-period discriminator (MPD) and multi-resolution discriminator (MRD), using a hinge formulation.
Under the conditions of a waveform sampling rate of 22.05 kHz and a spectral frame shift of approximately 11.6 ms, our proposed APNet2 vocoder significantly outperforms
the original APNet and Vocos, and is comparable to HiFi-GAN and iSTFTNet in terms of synthesized speech quality for both analysis-synthesis and TTS tasks.
In terms of efficiency, the inference speed of the proposed APNet2 is obviously faster than that of HiFi-GAN and iSTFTNet.

The rest of this paper is organized as follows. Section \ref{sec:rw} briefly reviews some related works including iSTFTNet, Vocos, and APNet.
In Section \ref{sec:pagestyle}, we give details of our proposed APNet2 vocoder.
The experimental results and analysis are presented in Section \ref{sec:exp}.
Finally, we make a conclusion and preview some areas of future research in Section \ref{sec:con}.

\section{Related Work}
\label{sec:rw}

\begin{figure}
	\includegraphics[width=\textwidth]{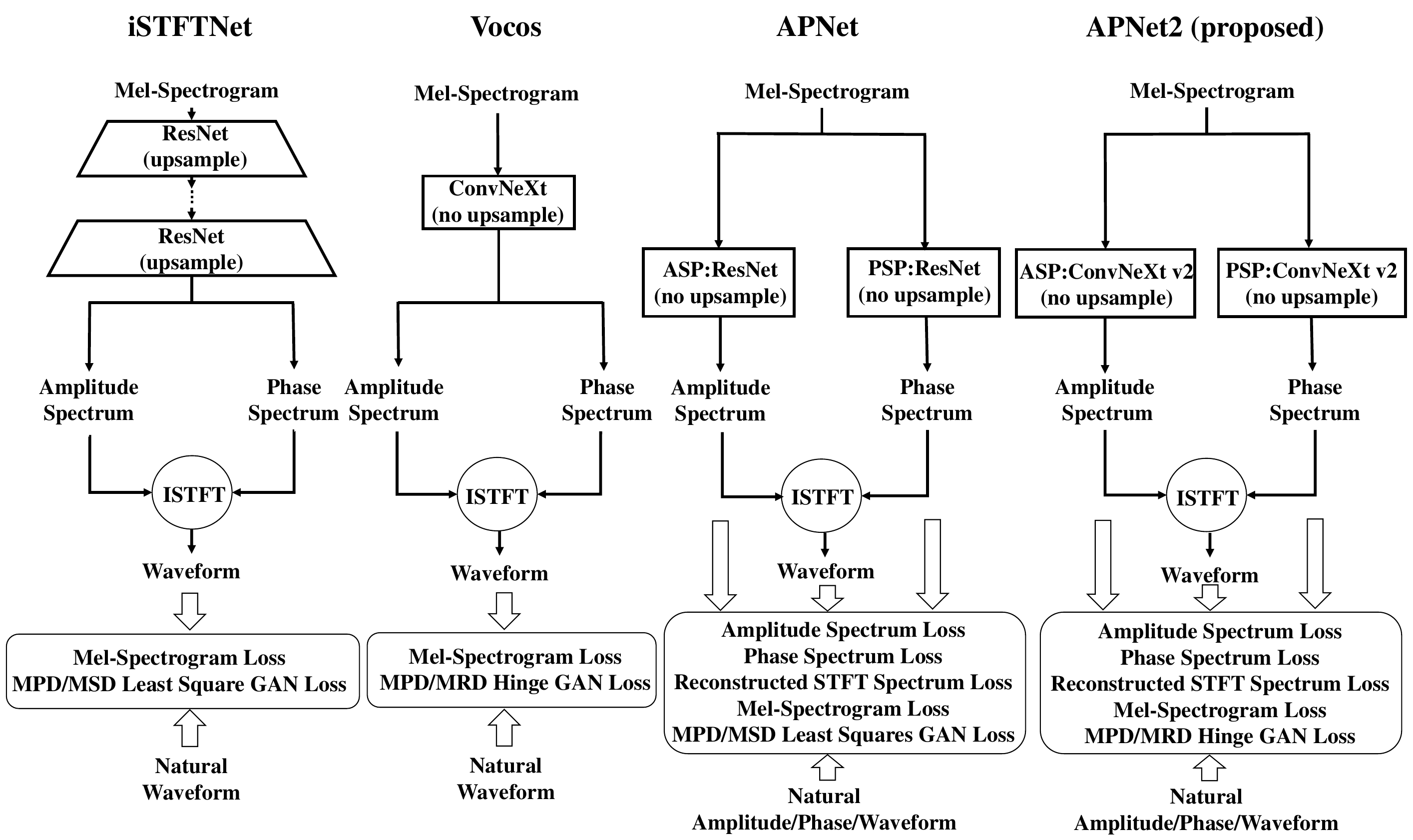}
	\caption{Overview of the concise architectures and loss functions of iSTFTNet, Vocos, APNet, and our proposed APNet2.} \label{fig1}
\end{figure}

Neural vocoders have garnered extensive attention in the fields of signal processing and machine learning.
The method of using ISTFT to avoid directly predicting time-domain waveforms has gained traction and demonstrated progress \cite{kaneko2022iSTFTNet,siuzdak2023vocos,ai2023apnet,oyamada2018generative,gritsenko2020spectral,neekhara2019expediting}.
Fig. \ref{fig1} shows the concise architectures and loss functions of representative models, including iSTFTNet \cite{kaneko2022iSTFTNet}, Vocos \cite{siuzdak2023vocos}, APNet \cite{ai2023apnet}, and the proposed APNet2.

\subsection{iSTFTNet}
As illustrated in Fig. \ref{fig1}, iSTFTNet \cite{kaneko2022iSTFTNet} initially processes the mel-spectrogram through multiple residual convolutional neural networks (ResNets) with upsampling operations to obtain the amplitude and phase spectra.
Subsequently, it reconstructs the waveform through the ISTFT operation.
It is evident that, through the upsampling layers, the amplitude and phase spectra predicted by iSTFTNet have higher temporal resolution and lower frequency resolution.
As mentioned in the original paper \cite{kaneko2022iSTFTNet}, upsampling operations are inevitable to ensure the quality of the synthesized speech.
Hence, iSTFTNet does not achieve a truly all-frame-level amplitude and phase prediction, leaving room for improvement in terms of inference efficiency.
The loss functions employed by iSTFTNet align with those of HiFi-GAN \cite{kong2020hifi}, which includes mel-spectrogram loss, feature matching loss, and MPD/multi-scale discriminator (MSD) based least squares GAN losses.
Experimental results show that the iSTFTNet achieves faster inference speed than HiFi-GAN, and its synthesized speech quality is comparable to that of HiFi-GAN.

\subsection{Vocos}
As illustrated in Fig. \ref{fig1}, Vocos \cite{siuzdak2023vocos} is an all-frame-level neural vocoder, which simultaneously predicts the amplitude and phase spectra at the original temporal resolution without any upsampling operations.
Vocos employs ConvNeXt block \cite{liu2022convnet} as the backbone network for better modeling capability compared to ResNet. 
The ConvNeXt block consists of a depth-wise convolution with a larger-than-usual kernel size, immediately followed by an inverted bottleneck.
This bottleneck projection raises the feature dimensionality through point-wise convolution, and during this process, Gaussian error linear unit (GELU) \cite{hendrycks2016gaussian} activation is utilized.
Normalization is employed between each block layer.
Then a multi-layer perceptron (MLP) layer downsamples the features to the original dimensionality. 
In terms of loss functions, Vocos has implemented several improvements compared to HiFi-GAN and iSTFTNet, including mel-spectrogram loss, feature matching loss, and MPD/MRD-based least squares GAN losses.
Empirical evidence confirms that Vocos achieves significantly faster inference speeds compared to HiFi-GAN while maintaining the quality of synthesized speech.
However, a common issue existing in both iSTFTNet and Vocos is the treatment of phase prediction as a black box without explicitly modeling it.
This approach may potentially impact the accuracy of phase prediction and, consequently, the quality of synthesized speech.

\subsection{APNet}

As illustrated in Fig. \ref{fig1}, APNet \cite{ai2023apnet} consists of an ASP and a PSP.
These two components work in parallel to predict the amplitude and phase spectra, which are then employed to reconstruct the waveform through ISTFT.
The backbone of both the ASP and PSP is the ResNet without any upsampling operations.
Specifically, the ResNet consists of several parallel residual convolution blocks (ResBlocks), featuring a large number of dilated convolutions.
Then, the outputs of each ResBlock are summed, averaged, and finally activated by a leaky rectified linear unit (ReLU) \cite{maas2013rectifier} activation.
In contrast to ASP, what distinguishes PSP is its emphasis on the characteristics of the wrapped phase and the introduction of a phase parallel estimation architecture at the output end.
The parallel estimation architecture is composed of two parallel linear convolutional layers and a phase calculation formula denoted as $\Phi$, simulating the process of computing phase spectra from short-time complex spectra.
The formula $\Phi$ is a bivariate function defined as follows:
\begin{equation}
	\Phi(R,I)=\arctan\left(\frac{I}{R}\right)-\frac{\pi}{2}\cdot Sgn^*(I)\cdot\left[Sgn^*(R)-1\right],
 	\label{eq1}
\end{equation}
where $R$ and $I$ represent the pseudo-real and imaginary parts output by the two parallel linear convolutional layers, respectively; $Sgn^*(x)$ is a new symbolic function defined in \cite{ai2023apnet}: when $x \geq 0$,  $Sgn^*(x)=1$, otherwise $Sgn^*(x)=-1$.

A series of loss functions are defined in APNet to guide the generation of spectra and waveforms, including:
1) amplitude spectrum loss $\mathcal L_A$, which is the $L^2$ distance of the predicted log amplitude spectrum and the natural one;
2) phase spectrum loss $\mathcal L_P$, which is the sum of instantaneous phase loss, group delay loss, and phase time difference loss, all activated by negative cosine anti-wrapping function, measuring the gap between the predicted phase spectrum and the natural one at various perspectives;
3) reconstructed STFT spectrum loss $\mathcal L_S$, which includes the STFT consistency loss between the reconstructed STFT spectrum and the consistent one, and $L^1$ distance of the real parts and imaginary parts between the reconstructed STFT spectrum and the natural one;
4) final waveforms loss $\mathcal L_W$, which is the same as used in HiFi-GAN \cite{kong2020hifi}, including mel-spectrogram loss, feature matching loss, and MSD/MPD least squares GAN loss.
We still use the names of these functions in APNet2, but only change the definition of some functions which will be introduced in section \ref{ssec:loss}.

\section{Proposed Method}
\label{sec:pagestyle}
\begin{figure}
	\centering
	\includegraphics[width=0.85\textwidth]{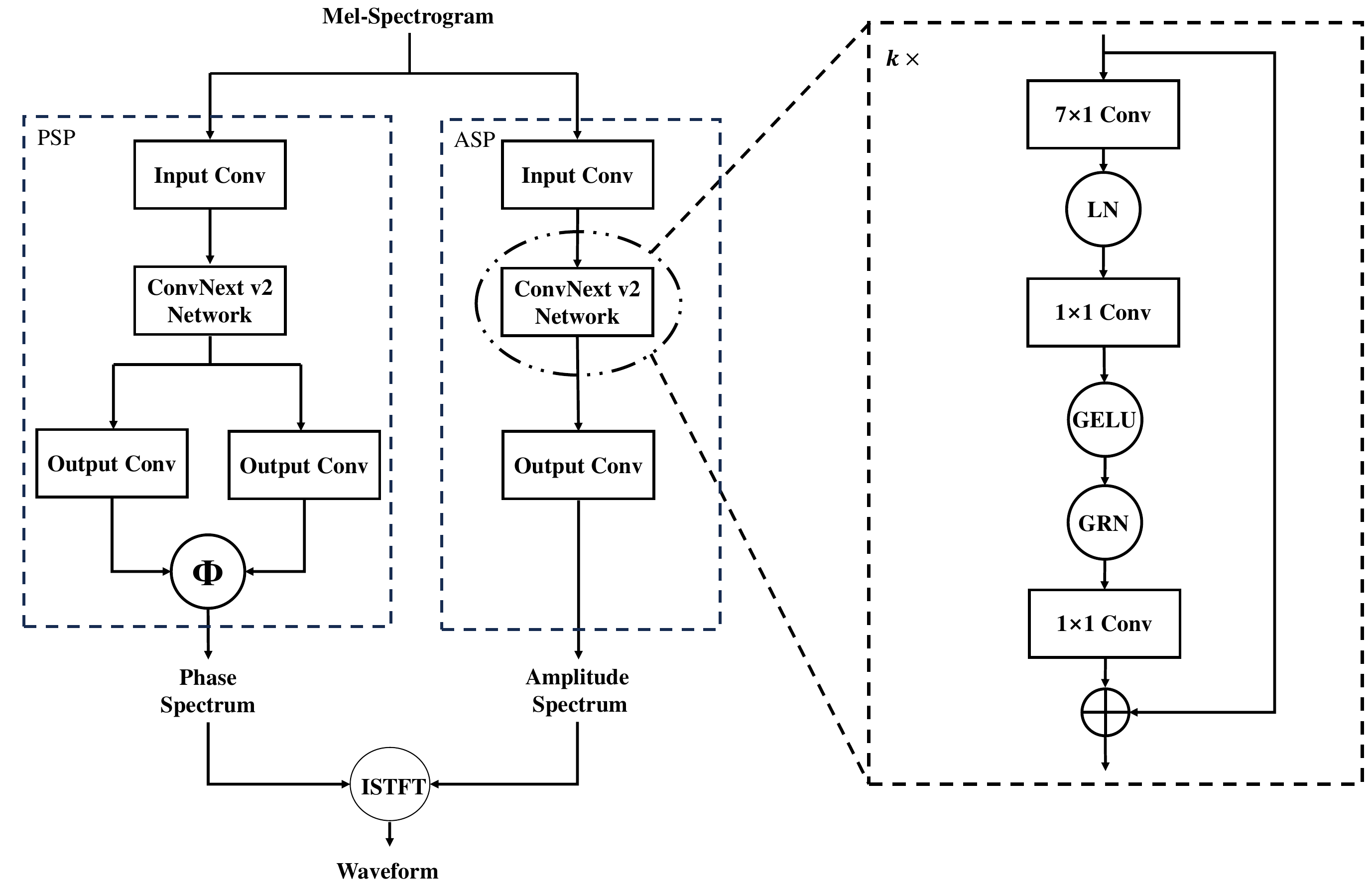}
	\caption{The architecture of APNet2. $\mathrm{\Phi}$ and ISTFT represent the phase calculation formula and inverse short-time Fourier transform, respectively. The content in the dotted box is the specific structure of the ConvNeXt v2 block, where LN, GELU, and GRN represent layer normalization, Gaussian error linear unit, and global response normalization, respectively. } \label{fig2}
\end{figure}

As illustrated in Fig. \ref{fig2}, the proposed APNet2 vocoder directly predicts speech amplitude spectra and
phase spectra at unique raw temporal resolution from input acoustic features (e.g., mel-spectrogram) in parallel.
Subsequently, the amplitude and phase spectra are reconstructed to the STFT spectrum, from which the waveform is ultimately recovered via ISTFT.

\subsection{Model Structure}
\label{ssec:block}
The proposed APNet2 vocoder comprises an ASP and a PSP.
The ASP and PSP aim to directly predict the logarithmic amplitude spectrum and the wrapped phase spectrum from the input mel-spectrogram, respectively.
The structure of ASP is a cascade of an input convolutional layer, a ConvNeXt v2 network \cite{woo2023convnext}, and an output convolutional layer, while that of PSP is a cascade of input convolutional layers, a ConvNeXt v2 network, and a phase parallel estimation architecture.
The phase parallel estimation architecture, adopted from APNet \cite{ai2023apnet}, is specially designed to ensure the direct output of the wrapped phase spectrum.

We employ ConvNeXt v2 as the backbone for both the ASP and PSP rather than the ResNet used in APNet because ConvNeXt v2 has demonstrated strong modeling capabilities in the field of image processing.
The ConvNeXt v2 is a cascade of $k$ ConvNeXt v2 blocks.
As shown in Fig. \ref{fig2}, each ConNeXt v2 block contains a large-kernel-sized depth-wise convolutional layer, a layer normalization operation, a point-wise convolutional layer that elevates feature dimensions, a GELU activation, a global response normalization (GRN) layer, and another point-wise convolutional layer that restores features to their original dimensionality.
Finally, residual connections are employed, with the input added to the output of the last point-wise convolutional layer as the block's output.
Compared to ConvNeXt \cite{liu2022convnet} used in Vocos, ConvNeXt v2 introduces a GRN layer positioned after the dimension-expansion MLP and drop layer scale operation, aiming to increase the contrast and selectivity of channels.
The GRN layer consists of global feature aggregation, feature normalization, and feature calibration, improving the representation quality by enhancing the feature diversity.

\subsection{Training Criteria}
\label{ssec:loss}

\begin{figure}
	\centering
	\includegraphics[width=0.7\textwidth]{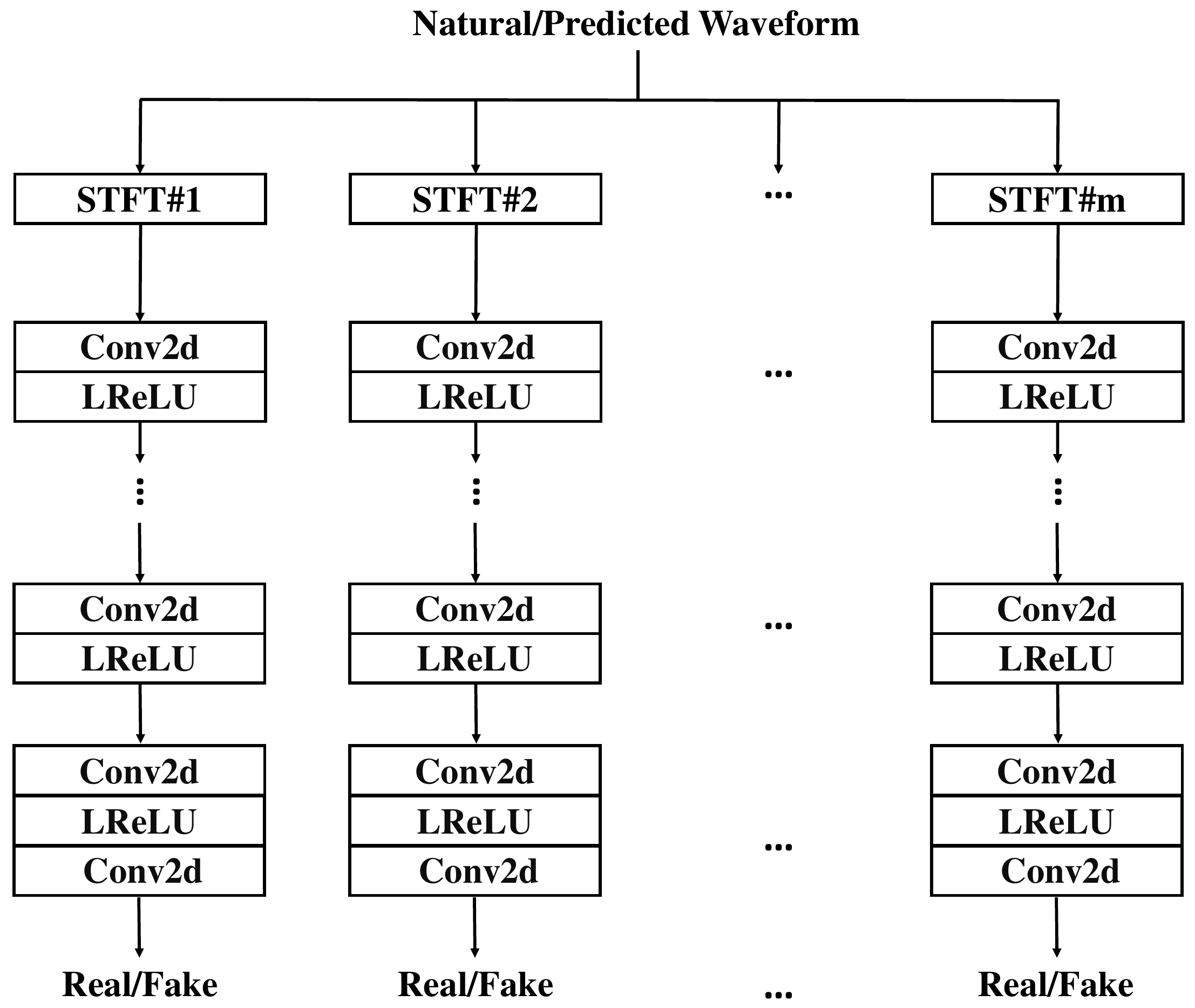}
	\caption{Architecture of MRD. STFT\#$m$ denotes computing the amplitude spectrum using the $m$-th STFT parameter set and LReLU denotes the leaky ReLU activation.} \label{fig3}
\end{figure}

In the APNet \cite{ai2023apnet}, we incorporate MPD and MSD as borrowed from \cite{kong2020hifi}.
However, the high parameter count of MSD results in slow training speed.
As mentioned in \cite{lee2023phaseaug}, the accuracy discriminative of MSD can reach 100\% during training, indicating overfitting of this discriminator.
In the proposed APNet2, we replaced MSD with MRD \cite{jang2021univnet}, whose key elements are strided 2-D convolutions and leaky ReLU activations  \cite{maas2013rectifier}, as shown in Fig. \ref{fig3}.
The MRD is divided into multiple sub-discriminators, each operating at different resolutions.
Each sub-discriminator first extracts amplitude spectra at a certain temporal and spectral resolution from input natural/predicted waveforms with certain STFT parameters, and then outputs a discriminant value.
The MRD is expected to cover a wide range of scales of speech waveforms as comprehensively as possible.
As recommended in \cite{zeghidour2021soundstream}, we discard the least squares GAN loss used in the APNet, and adopt a hinge GAN loss here as follows:

\begin{equation}
	\mathcal L_{GAN-G}(\bm{\mathit{\hat{x}}})=\frac{1}{L}\sum_l \max(0,1-D_l(\bm{\mathit{\hat{x}}} )),
 	\label{eq2}
\end{equation}
\begin{equation}
	\mathcal L_{GAN-D}(\bm{\mathit{x} } ,\bm{\mathit{\hat{x}}} )=\frac{1}{L}\sum_l \max(0,1-D_l(\bm{\mathit{x} } ))+\max(0,1+D_l(\bm{\mathit{\hat{x}}} )),
 	\label{eq3}
\end{equation}
where $D_l$ is the $l$-th sub-discriminator in MPD and MRD, and $L$ is the total number of sub-discriminators in MPD and MRD.
$\bm{x}$ and $\bm{\mathit{\hat{x}}} $ are the natural and the synthesized waveforms, respectively.

We retain all the losses in the APNet \cite{ai2023apnet} which are defined on amplitude spectra, phase spectra, reconstructed STFT spectra, and final waveforms.
Refer to \cite{ai2023neural}, we make slight modifications to the phase-related losses, i.e., using a linear function $f_{AW}$ as the anti-wrapping function and discarding the negative cosine function used in the APNet.
In our preliminary experiments, we confirmed that the linear form is more suitable for phase prediction than the cosine form.
The linear function $f_{AW}$ is defined as:
\begin{equation}
	f_{AW}(x)=\left|x-2\pi\cdot round\left(\frac{x}{2\pi}\right)\right|,
	\label{eq4}
\end{equation}
and used to activate the instantaneous phase error, group delay error, and phase time difference error between predicted and natural phases for mitigating the error expansion issue caused by phase wrapping.

We use the generative adversarial strategy to train APNet2, and the generator loss is the linear combination of the loss functions mentioned above:
\begin{equation}
	\mathcal L_G=\lambda_A\mathcal L_A+\lambda_P\mathcal L_P+\lambda_S\mathcal L_S+\lambda_W\mathcal L_W,
	\label{eq5}
\end{equation}
where $\lambda_A$, $\lambda_P$, $\lambda_S$, and $\lambda_W$ are hyperparameters, taking the same values as in APNet.
$\mathcal L_A$, $\mathcal L_P$, $\mathcal L_S$, and $\mathcal L_W$ are amplitude spectrum loss, phase spectrum loss, reconstructed STFT spectrum loss, and final waveforms loss, respectively, where $\mathcal L_W$ includes MPD/MRD hinge GAN loss $\mathcal L_{GAN-G}$ defined as Eq. \ref{eq2}, feature matching loss, and mel-spectrogram loss.
The discriminator loss is $\mathcal L_D = \mathcal L_{GAN-D}$.

\section{Experiments}
\label{sec:exp}

\subsection{Experimental Setup}
\label{ssec:expset}
\subsubsection{Dataset.} We used the LJSpeech \cite{ito2017lj} dataset for our experiments, which consists of 13,100 audio clips of a single English female speaker and is of about 24 hours.
The audio sampling rate is 22.05 kHz with a format of 16-bit PCM.
We randomly selected 12,000 audio clips for training, 100 for validation, and 500 for testing.
Spectral features (e.g., 80-dimensional mel-spectrograms, amplitude spectra, phase spectra) were extracted by STFT with an FFT point number of 1024, frame shift of 256 (i.e., approximately 11.6 ms), and frame length of 1024 (i.e., approximately 46.4 ms).
Note that the mel-spectrogram is not a full-band spectrum.
It only covers the frequency range of 0 to 8000 Hz to be aligned with the common configuration \cite{kong2020hifi,kaneko2022iSTFTNet}.
Therefore, the vocoders generate waveforms from the input mel-spectrogram in this configuration that, in fact, imply the operation of BWE.

\subsubsection{Implementation.}\label{above}
In the proposed APNet2 vocoder, the number of ConvNeXt v2 blocks $k$ in both ASP and PSP was set to 8.
In each ConvNeXt v2 block, the kernel size and channel size of the large-kernel-sized depth-wise convolutional layer were set to 7 and 512, respectively.
The channel sizes of the first and the last 1$\times$1 point-wise convolutional layers were set to 512 and 1536, respectively.

We trained our proposed APNet2 vocoder up to 2 million steps, with 1 million steps per generator and discriminator, on a single Nvidia 2080Ti GPU.
During training, we randomly cropped the audio clips to 8192 samples and set the batch size to 16.
The model is optimized using the AdamW optimizer \cite{loshchilov2018decoupled} with $\beta_1=0.8$, $\beta_2 = 0.99$, and weight decay of 0.01. The learning rate was set initially to $2 \times 10^{-4}$ and scheduled to decay with a factor of 0.999 at every epoch.

\subsubsection{Baselines.} We compared the proposed model APNet2 to HiFi-GAN\footnote{https://github.com/jik876/hifi-gan}, iSTFTNet\footnote{https://github.com/rishikksh20/iSTFTNet-pytorch}, Vocos\footnote{https://github.com/charactr-platform/vocos} and APNet\footnote{https://github.com/yangai520/APNet}. 
All these compared vocoders were trained on the same settings as mentioned in APNet2 using their open-source implementations.

\subsubsection{Tasks.}
We applied our proposed APNet2 and baseline vocoders to two tasks in our experiments, i.e., the analysis-synthesis task and the TTS task, which employed natural and predicted mel-spectrograms as vocoders' input, respectively.
For the TTS task, we used a Fastspeech2-based acoustic model\footnote{https://github.com/ming024/FastSpeech2} \cite{ren2020fastspeech} to predict the mel-spectrograms from texts.

\subsection{Evaluation}
\label{ssec:eva}
We use both objective and subjective evaluations to compare the performance of these vocoders.
Five objective metrics for evaluating the quality of synthesized speech used in our previous work \cite{ai2023apnet} were adopted here, including the signal-to-noise ratio (SNR), root mean square error (RMSE) of log amplitude spectra (denoted by LAS-RMSE), mel-cepstrum distortion (MCD), RMSE of F0 (denoted by F0-RMSE), and V/UV error. 
To evaluate the inference efficiency, the real-time factor (RTF), which is defined as the ratio between the time consumed to generate speech waveforms and their total duration, was also utilized as an objective metric. 
In our implementation, the RTF value was calculated as the ratio between the time consumed to generate all 500 test sentences using a single Nvidia 2080Ti GPU or a single Intel Xeon E5-2620 CPU core and the total duration of the test set.

To assess the subjective quality, we conducted mean opinion score (MOS) tests to compare the naturalness of these vocoders. Each MOS test involved twenty test utterances synthesized by different vocoders, alongside natural utterances. We gathered feedback from a minimum of 25 native English listeners on the Amazon Mechanical Turk crowdsourcing platform\footnote{https://www.mturk.com}. Listeners were asked to rate the naturalness on a scale of 1 to 5, with a score interval of 0.5.

\begin{table}
	\centering
	\caption{Objective results of HiFi-GAN, iSTFTNet, Vocos, APNet, and the proposed APNet2 on the test set of the LJSpeech dataset for the analysis-synthesis task. Here, "$a\times$" represents $a\times$ real time.}\label{tab1}
	\adjustbox{width=\textwidth}{
		\begin{tabular}{l c c c c c c c}
			\hline
			\hline
			\multirow{2}{*} & {\scriptsize SNR} & {\scriptsize LAS-RMSE}& {\scriptsize MCD} &{\scriptsize F0-RMSE} & {\scriptsize V/UV error}& {\scriptsize RTF}& {\scriptsize RTF}\\ &  {  {\scriptsize(dB)$\uparrow$} }& { {\scriptsize(dB)$\downarrow$}}&{ {\scriptsize(dB)$\downarrow$}} &{ {\scriptsize (cent)$\downarrow$}}&{\scriptsize (\%)$\downarrow$}&{\scriptsize (GPU)$\downarrow$}&{\scriptsize (CPU)$\downarrow$}\\
			\hline
			{\scriptsize HiFi-GAN} & \textbf{3.93}&6.44&\textbf{1.62}&45.08&\textbf{5.20}&0.0296 (33.75$\times$) & 0.297 (3.36$\times$)\\
			{\scriptsize iSTFTNet} & 3.83&6.78&1.71&\textbf{43.60}&5.28&0.0057 (175.95$\times$)&0.148 (6.76$\times$)\\
			{\scriptsize Vocos} & 2.30&6.97&2.31&148.33&10.64&\textbf{0.0012 (869.90$\times$)}&\textbf{0.009(116.45$\times$)}\\
			{\scriptsize APNet} & 2.65&7.03&2.07&46.68&5.53&0.0028 (358.17$\times$)&0.039(25.84$\times$)\\
			\hline
			{\scriptsize \textbf{APNet2}} & 3.84&\textbf{6.34}&1.73&44.33&5.31&0.0015 (665.05$\times$)&0.021(47.73$\times$)\\
			
			\hline
			\hline
	\end{tabular}}
\end{table}

\noindent The objective results on the test sets of the LJSpeech dataset for the analysis-synthesis task are listed in Table \ref{tab1}.
HiFi-GAN obtained the highest scores for most objective metrics. 
Our proposed APNet2 was comparable to HiFi-GAN and iSTFTNet, and outperformed the Vocos and APNet. 
In terms of RTF, the inference speed of APNet2 was second only to the fastest Vocos, and significantly faster than other vocoders on both GPU and CPU. 
However, Vocos exhibited a noticeable disadvantage in all objective metrics, particularly with a high F0-RMSE, indicating noticeable pronunciation errors.
Furthermore, ISTFT-based vocoders (i.e., iSTFTNet, Vocos, APNet, and APNet2) significantly improved inference speed compared to HiFi-GAN. 
This also validated the effectiveness of predicting low-resolution spectra rather than high-resolution time-domain waveforms for efficiency enhancement. 

The subjective MOS test results on the test set of the LJSpeech dataset for both analysis-synthesis and TTS tasks are listed in Table. \ref{tab2}. 
For the analysis-synthesis task, iSTFTNet got the highest average MOS among all vocoders. 
The average MOS of the proposed APNet2 vocoder was lower than that of iSTFTNet and HiFi-GAN. 
To verify the significance of the differences between paired vocoders, we calculated the $p$-value of a $t$-test for paired MOS sequences of the test set. 
Interestingly, the $p$-value of the results between iSTFTNet and APNet2 was 0.148, and the $p$-value of the results between HiFi-GAN and APNet2 was 0.515. 
This indicates that, in terms of synthesized speech quality, there was no significant difference between the proposed APNet2 and two vocoders with high average MOS (i.e., iSTFTNet and HiFi-GAN).
Moreover, APNet2 was significantly ($p < 0.01$) better than Vocos and APNet regarding the MOS scores. 
The conclusions in the TTS task were closely aligned with those in the analysis-synthesis task, and it is gratifying that in the TTS task, APNet2 achieved one of the highest average MOS scores. 
This indicates that APNet2 exhibited strong robustness when dealing with non-natural acoustic features. 
Furthermore, by observing the objective and subjective results between APNet and APNet2, it can be concluded that the introduction of ConvNeXt v2, MRD, etc., effectively improved the model's modeling capabilities, making it suitable for waveform generation in high waveform sampling rates and long spectral frame shift scenarios. 
For a more intuitive experience, please visit our demo page\footnote{Source codes are available at https://github.com/redmist328/APNet2. Examples of
	generated speech can be found at https://redmist328.github.io/APNet2\_demo}.

\begin{table}
	\centering
	\caption{Subjective results of HiFi-GAN, iSTFTNet, Vocos, APNet, and the proposed APNet2 on the test set of the LJSpeech dataset. Here, "AS" and "TTS" represent analysis-synthesis and TTS tasks, respectively.}\label{tab2}
	 \adjustbox{width=0.7\textwidth}{
	\begin{tabular}{l c c c c}
		\hline
		\hline
		 &\quad\quad\quad &{ {\scriptsize { MOS (AS)}}} &\quad\quad\quad &{ {\scriptsize { MOS (TTS)}}}\\
		\hline
		{ Natural Speech}&\quad\quad\quad &4.02$\pm$0.108&\quad\quad\quad &3.93$\pm$0.176\\
		
		\hline
		{ {\scriptsize HiFi-GAN}}&\quad\quad\quad &  \textbf{3.88$\pm$0.153}&\quad\quad\quad &\textbf{3.58$\pm$0.259}\\
		{ {\scriptsize iSTFTNet}}&\quad\quad\quad & \textbf{3.93$\pm$0.148}&\quad\quad\quad &\textbf{3.66$\pm$0.232}\\
		{ {\scriptsize Vocos}} &\quad\quad\quad &3.50$\pm$0.296&\quad\quad\quad &3.25$\pm$0.379\\
		{ {\scriptsize APNet}} & \quad\quad\quad&3.53$\pm$0.256 &\quad\quad\quad &3.39$\pm$0.329\\
		\hline
		{ {\scriptsize \textbf{APNet2}}} & &\textbf{3.83$\pm$0.296}& &\textbf{3.66$\pm$0.230}\\
		\hline
		\hline
	\end{tabular}}

\end{table}

\subsection{Analysis and discussion}
\label{ssec:ana}

We did some analysis experiments and ablation studies to examine the effectiveness of each key component in the APNet2 vocoder. 
We only make objective evaluations here to compare different vocoder variations.

As shown in Table~\ref{tab1} and Table~\ref{tab2}, the performance of Vocos was disappointing. 
However, in the original paper of Vocos \cite{siuzdak2023vocos}, the authors used the 100-dimensional full-band log-mel-spectrogram as input and achieved perfect results. 
Therefore, we replicated the original configuration in \cite{siuzdak2023vocos} and built Vocos w 100-dim-mel. 
As shown in the results in Table \ref{tab3}, when using 100-dimensional full-band log-mel-spectrograms as input, Vocos exhibited significant improvement in all objective metrics. 
This suggests that Vocos was highly sensitive to the frequency band range of input features, making it difficult to implicitly achieve BWE. 
This could be the reason for Vocos' poor performance when using 80-dimensional narrow-band mel-spectrograms as input. 

\begin{table}
	\centering
	\caption{Objective results of different variants of APNet2 and Vocos for the analysis-synthesis task.}\label{tab3}
	\adjustbox{width=\textwidth}{
		\begin{tabular}{l c c c c c }
			\hline
			\hline
			\multirow{2}{*} & {\scriptsize SNR} & {\scriptsize LAS-RMSE}& {\scriptsize MCD} &{\scriptsize F0-RMSE} & {\scriptsize V/UV error}\\ &  {  {\scriptsize(dB)$\uparrow$} }& { {\scriptsize(dB)$\downarrow$}}&{ {\scriptsize(dB)$\downarrow$}} &{ {\scriptsize (cent)$\downarrow$}}&{\scriptsize (\%)$\downarrow$}\\
			\hline
			{\scriptsize Vocos } & 2.30&6.97&2.31&148.33&10.64 \\
			{\scriptsize  Vocos w/ 100-dim-mel } & 2.96&5.84&1.80&56.59&6.81 \\
			\hline
			{\scriptsize APNet2} & \textbf{3.84}&6.34&1.73&44.33&5.31\\
			{\scriptsize APNet2 w/ 100-dim-mel }&3.42&\textbf{5.50}&\textbf{1.64}&48.69&5.88\\
			{\scriptsize APNet2 w/o ConvNeXt v2} &1.95&6.90&2.36&76.40&8.18\\
			{\scriptsize APNet2 w/o MRD } &1.98&7.14&2.48&65.69&8.25\\
			{\scriptsize APNet2 w/o HingeGAN}  &2.61&6.57&2.00&\textbf{35.69}&\textbf{4.64} \\
			\hline
			\hline
		\end{tabular}
	}
\end{table}

\noindent Subsequently, we compared APNet2 with some of its variants. 
The results are also listed in Table~\ref{tab3}.
1) Firstly, we also used 100-dimensional full-band log-mel-spectrogram as APNet2's input (i.e., APNet2 w/ 100-dim-mel). 
Compared to APNet2, APNet2 with 100-dim-mel showed significant improvements in the amplitude-related metrics (i.e., LAS-RMSE and MCD). 
It is reasonable because 100-dimensional mel-spectrograms provide complete high-frequency amplitude information. 
2) Then, we replaced the ConvNeXt v2 with the original ResNet and built APNet2 w/o ConvNeXt v2. 
Clearly, with the absence of ConvNeXt v2, all objective metrics show a sharp decline, confirming the superiority of ConvNeXt v2 in modeling capability. 
3) Thirdly, in the GAN-based losses, MRD was replaced with MSD (i.e., APNet2 w/o MRD). 
The performance of APNet2 w/o MRD also significantly deteriorated. 
This indicates that MRD was better suited for waveform discrimination. 
4) Finally, also in the GAN-based losses, we adopted the original least squares form rather than the hinge form (i.e., APNet2 w/o HingeGAN). 
Although using the least squares significantly improved the accuracy of F0, the waveform and spectral-related metrics were significantly degraded. 
This suggests that employing the hinge GAN loss was beneficial in improving overall waveform and spectral quality to some extent.

\section{Conclusions}
\label{sec:con}
In this paper, we proposed a novel APNet2 vocoder, addressing the performance limitations of the original APNet that were constrained by the waveform sampling rate and spectral frame shift.
In comparison to APNet, the improvements in APNet2 included using ConvNeXt v2 as the backbone network for amplitude and phase predictions, introducing MRD into the GAN-based losses, and employing the hinge GAN form, etc. 
Experimental results demonstrated that APNet2 can achieve high-quality and efficient waveform generation at the configuration with a waveform sampling rate of 22.05 kHz and spectral frame shift of approximately 11.6 ms.
Moreover, ablation studies verified the effectiveness of the key components in the APNet2 vocoder.
Further applying the APNet2 vocoder to other speech generation tasks (e.g., SE and VC) will be the focus of our future work. 
%
%
%
%
\bibliographystyle{splncs04_unsort}
\bibliography{mybibliography}

\end{document}